\documentclass[10pt,a4paper,english]{article}
\pdfoutput=1
\usepackage[ansinew]{inputenc}
\usepackage{babel}
\usepackage{latexsym}
\usepackage{amsmath}
\usepackage{amsfonts}
\usepackage{amssymb}
\usepackage{textcomp}
\usepackage[pdftex]{hyperref}
\usepackage{graphicx}
\usepackage{enumerate}
\usepackage{float} %imagenes donde se quiera
\usepackage{pdfpages} %incluir pdf externo

\usepackage{cancel}

\usepackage[a4paper,centering,vcentering,marginratio=1:1,hscale=0.85,vscale=0.81]{geometry}
\numberwithin{equation}{section}

\usepackage{fancyhdr}
\fancypagestyle{plain}{
\lhead{}
\chead{}
\rhead{\bfseries Entanglement and the infrared structure of QED}
\lfoot{Master in Fundamental Physics  07-08}
\rfoot{Eduardo Martín Martínez}
\cfoot{}

}

\makeatletter \def\cleardoublepage{\clearpage\if@twoside \ifodd\c@page\else
   \thispagestyle{empty}
   \newpage
   \if@twocolumn\hbox{}\newpage\fi\fi\fi}
\makeatother

\def\openone{\leavevmode\hbox{\small$1$\normalsize\kern-.33em$1$}}

\newcommand{\ket}[1]{\left| {#1} \right\rangle}
\newcommand{\bra}[1]{\left\langle {#1} \right|}
	
\newcommand{\braket}[2]{\left\langle {#1}\left|{#2}\right.\right\rangle}
\newcommand{\proj}[2]{\left| {#1} \right\rangle\!\left\langle {#2} \right|}

\def\slashchar#1{\setbox0=\hbox{$#1$} % set a box for #1
\dimen0=\wd0 % and get its size
\setbox1=\hbox{/} \dimen1=\wd1 % get size of /
\ifdim\dimen0>\dimen1 % #1 is bigger
\rlap{\hbox to \dimen0{\hfil/\hfil}} % so center / in box
#1 % and print #1
\else % / is bigger
\rlap{\hbox to \dimen1{\hfil$#1$\hfil}} % so center #1
/ % and print /
\fi}

\newcommand{\cnm}[2]{\left[{#1},{#2}\right]}

\title{Entregables Información Cuántica}
\author{Eduardo Martín Martínez}
\date{}

\begin{document}
\includepdf{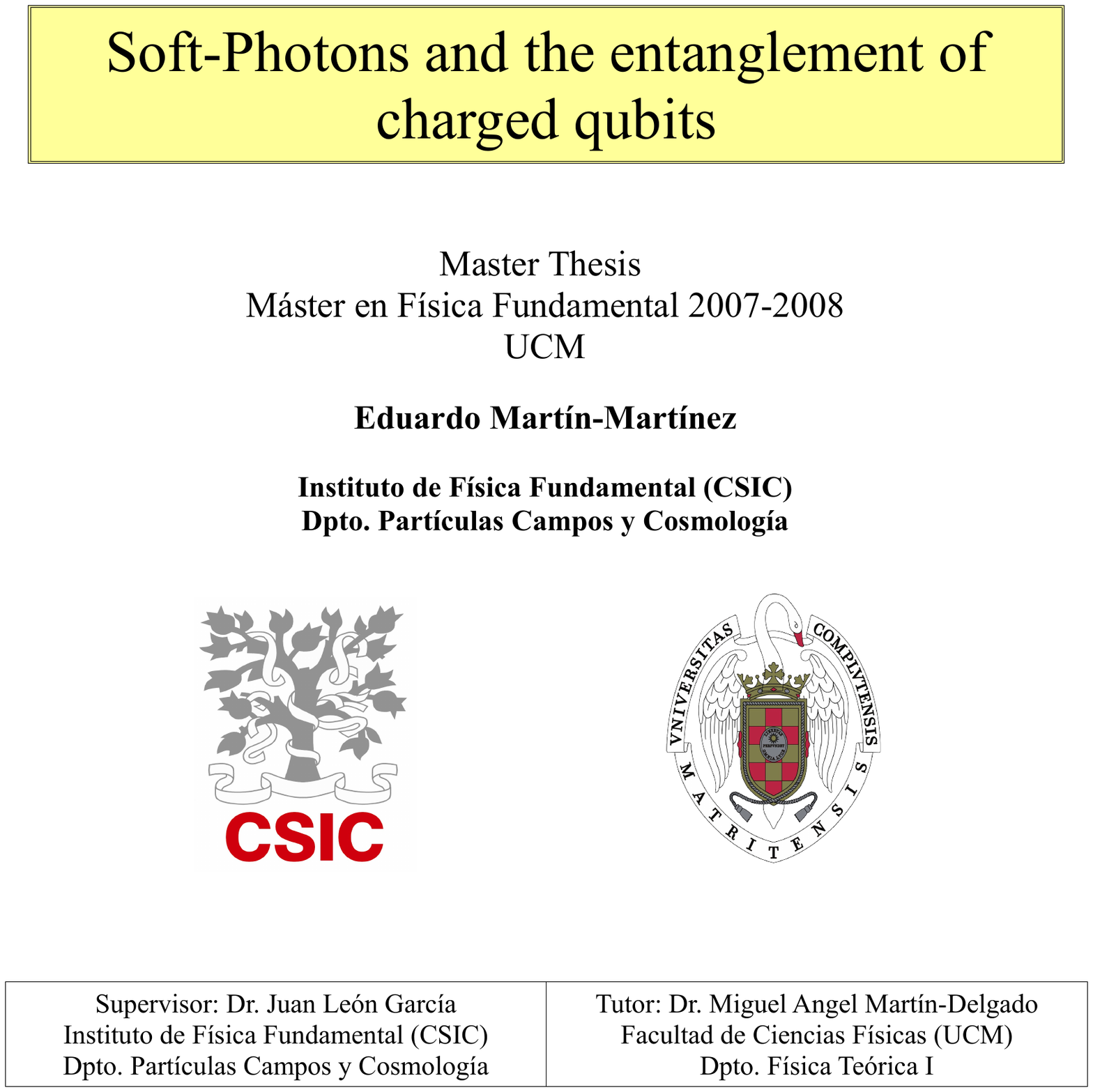}
\includepdf{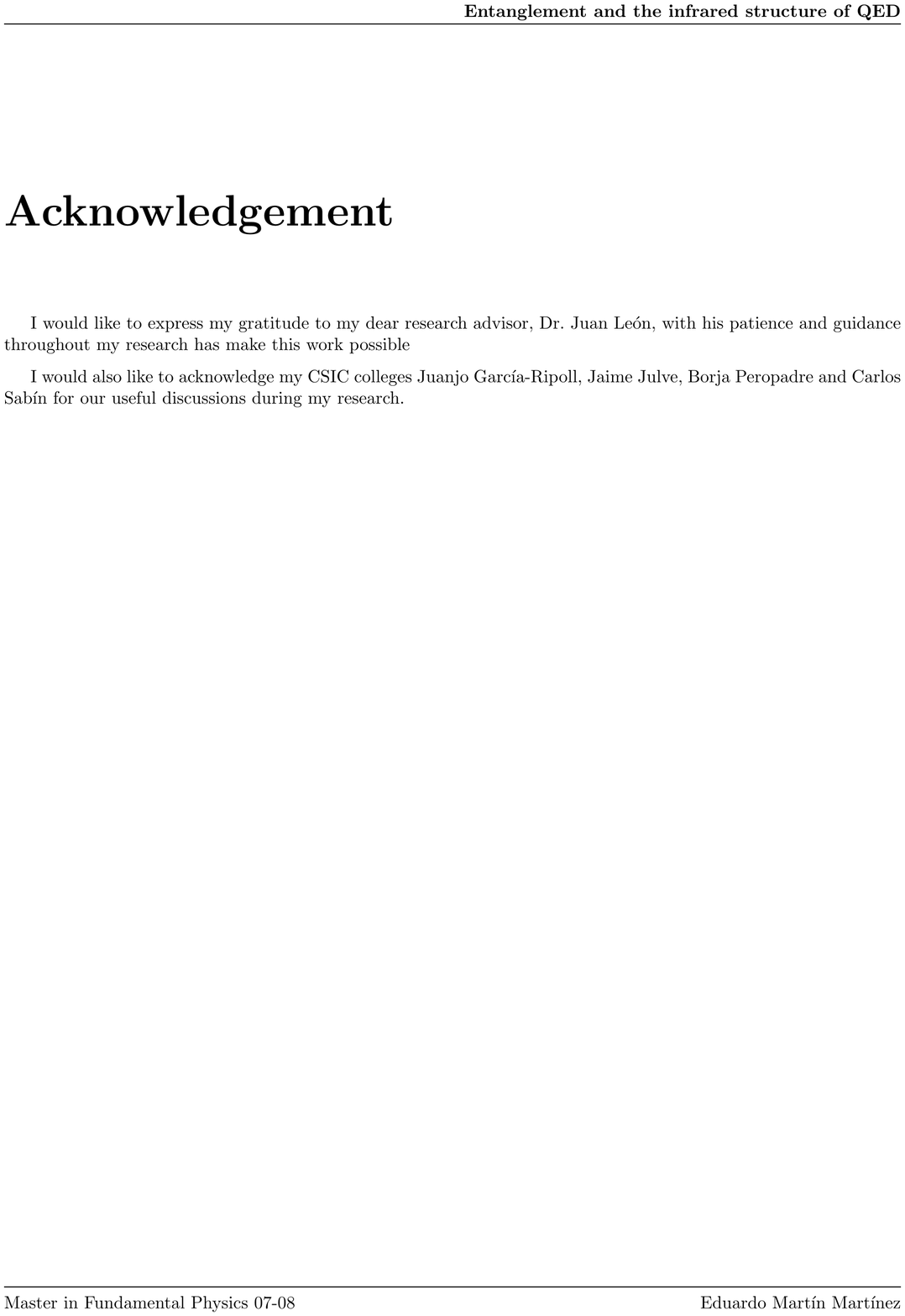}
\setcounter{page}{0}
\thispagestyle{plain}
\tableofcontents
\newpage

\section{Introduction}

In this work we analyse for the first time the effect of the infrared structure of QED on the entanglement of charged qubits.

The qubits used in Quantum Information Theory are built from selected degrees of freedom which belong to (more or less complex) quantum systems. One of the simplest cases could be the spin of a charged particle or the photons polarisation. A new discipline is born in order to properly analyse these systems, it is what Peres and Terno \cite{b1} called Relativistic Quantum Information Theory, which concerns the role of relativistic-like considerations in the Quantum Information Theory.

The attempts to include the dynamical effect of Quantum Electrodynamics (QED) on the correlations between particles are very recent. We might consider that they started with the studies of the effect of the QED spin-spin interactions on the entanglement \cite{b02} and the violation of Bell Inequalities due to QED \cite{b03}. Complementarily, an ``infinite'' entanglement generation between interacting systems momenta has been predicted \cite{b04}, as well as entanglement transfer between the space degrees of freedom and the qubits. Both cases in the QED frame.

However, the formulation of Quantum Information Theory in the relativistic context is still far from being complete. For instance, there is no satisfactory notion of localization \cite{blocaliz}, \cite{blocaliz2}, \cite{blocaliz3} or, for example, it is not clear what would be the proper spin operator in order to characterize that degree of freedom in the context of Relativistic Quantum Information Theory \cite{bcaban}.

In the last decade, starting from the work of Czachor \cite{b05} and, afterwards, a lot of papers such as \cite{b04}, \cite{b06}, \cite{b1}, \cite{b08}, \cite{b09}, \cite{b010}, \cite{b011},\cite{bTerno} and many others have considered relativistic effects on Quantum Information Theory, proposing a new phenomenology regarding the entanglement from a Quantum Field Theory approach.

One of the main problems of this kind of works is the need to use perturbation theory so as to characterize interacting and space-like separated qubits. The simplest way to do it is to describe them by means of free asymptotic states (``in'' and ``out'') as it is common in Scattering Theory, postulating that in the large time limit the interaction is adiabatically deactivated. With this assumption, the free field operators in the interaction picture would give us the description of physical particle states that properly evolve in the asymptotic regime.

Nevertheless, it is well known \cite{b2} that the free Hamiltonian is not the large $t$ limit of the QED Hamiltonian, therefore it does not make sense to identify the free states $b^\dagger_{\sigma}(p)\ket0$ (where $b^\dagger_{\sigma}(p)$ is in the interaction picture) with asymptotic states within QED. Complementarily, divergences appear in every order of the perturbation theory because of emission and absorption of soft photons (photons whose energy and momentum are much lesser than the characteristic masses and energies of the processes). However, the divergences that come from real soft photons are canceled by the divergences that come from virtual soft photons so that the cross sections are finite \cite{bWein}.

On the other hand, the states that evolve properly in the asymptotic regime are safe from infrared divergences \cite{b2} although, as it will be shown, the evolution under the electromagnetic interaction in the asymptotic limit makes it very difficult to identify charged particle states, actually, an undetermined number of soft photons appear, spoiling the gauge-invariance of the asymptotic states. What is more, in some papers it is said that the very concept of charged physical particle may not make sense rigourously speaking. This fact would be disastrous when you are trying to build qubits from charged particles.

This fact is directly related with the masslessness of the photon, it leads to a series of pathologies in QED whose effect on the qubits and their entanglement is still unknown: isolated poles (associated with the mass of the charged particles) do not appear in the energy spectrum of QED; as a matter of fact, corresponding to each one, there appears a threshold of continuous states corresponding to states with an arbitrary number of soft photons. In those conditions, the electron propagator no longer corresponds to a simple particle pole $(p^2-m^2)^{-1}$, instead, the exponent acquires an irrational value (which depends on the fine structure constant) that leads to a cut in the complex plane (concept of infraparticles \cite{bSchroer}.)

Along the second half of the XX century it has been discussed whether physical states could be associated to charged particles or not, and, if they could, what would be their nature. The first contributions to this topic come from Dirac himself \cite{b3}.

The effect of the infrared structure of QED on the entanglement of a system of two charged particles (and which , therefore, undergo electromagnetic interaction) has not been considered yet, although Peres and Terno \cite{b1} identified it as a problem, noticing that the concept of charged qubit is merely an idealization, and then, the properties of an entangled state of charged particles would be modified in an unknown way due to the presence of an undeterminate number of soft photons.

Recent works \cite{b4}, \cite{b5} have gone further into the procedure called ``dressing'': construction of asymptotic states with good enough properties to be identified with physical charged particles by means of restoring their gauge invariance. However, no one of the previous works have treated in depth the phase effects and divergences which appear derived from the asymptotic evolution \cite{b2}, \cite{b1} when the asymptotic states are built. In this work these topics are dealt with.

The objective of this work is, therefore, to characterize the effect of soft photons on the entanglement of two charged qubits, something that has not been done yet. In order to achieve it, we will need to extend the theory associated with the construction of asymptotic states and the ``dressing'' of these states so as to apply those results to our case. Here, we will compute for the first time the effect of the infrared structure of QED on the entanglement. Eventually, we will prove that all the correlation functions of the spin degrees of freedom (and particularly, EPR correlations) are not modified at all by the presence of soft photons when we are considering charged qubits built from proper asymptotic particle states.

In section 2, we raise the most general two charged particles state in the free case. In section 3 all the aspects concerning asymptotic dynamics (and the construction of asymptotic states) are treated. In section 4 we apply the ``dressing'' procedure in order to build physical states (gauge invariant) from asymptotic charged particles states. Eventually, in section 5 the entanglement of these physical asymptotic states is compared with the entanglement in the free case. Conclusions are summarized in section 6.

\section{Two Dirac's Qubits. Free case}

Let us start with the description of two charged qubits. We are only concerned with their spin degrees of freedom. As charged particle we will understand (in this section) the idealized concept of free particle \cite{b1} which is commonly used in the frame of Relativistic Quantum Information Theory.

The most general state of 2 free fermionic qubits may be written as\footnote{We could regard arbitrary spinorial or tensorial structures by simply generalizing the function $\varphi(p_1,p_2,\sigma_1,\sigma_2)$. There is no loss of generality with the choice made here, as we will see later.}
\begin{equation}\label{estadogeneral}\ket\Psi=\sum_{\sigma_1,\sigma_2}\int d^3p_1\,d^3p_2\, \varphi_{\sigma_1\sigma_2}(p_1,p_2)\,\ket{\sigma_1,p_1\,;\,\sigma_2,p_2}\end{equation}
Where we have defined
\begin{equation}\label{notacion}\ket{\sigma_1,p_1\,;\,\sigma_2,p_2}\equiv \ket{\sigma_1,\sigma_2}\ket{p_1,p_2}\equiv b^\dagger_{\sigma_2}(p_2)b^\dagger_{\sigma_1}(p_1)\ket0\end{equation}
Being $b^\dagger_{\sigma}(p)$ a fermionic operator\footnote{NB: In order to simplify notation we write the state $\ket{p_1,p_2}$ instead of $\ket{\vec p_1,\vec p_2}$} in the interaction picture. It must be satisfied that
\[\sum_{\sigma_1,\sigma_2}\int d^3p_1\, d^3p_2\, |\varphi_{\sigma_1\sigma_2}(p_1,p_2)|^2<\infty\]
Let us build the density matrix associated with this general state
\begin{equation}\label{ecspin}
\proj{\Psi}{\Psi}=\int d^3p_1\,d^3p_2\,d^3p'_1\,d^3p'_2\, S_{\sigma_1\sigma_2,\sigma_1'\sigma_2'}(p_1,p_2,p'_1,p'_2) \ket{p_1,p_2}\bra{p'_1,p'_2}
\end{equation}
where the $(4\times4)$ matrix $S_{\sigma_1\sigma_2,\sigma_1'\sigma_2'}(p_1,p_2,p'_1,p'_2)$ is given by
\begin{equation}\label{matriz}
S_{\sigma_1\sigma_2,\sigma_1'\sigma_2'}(p_1,p_2,p'_1,p'_2)=\varphi_{\sigma_1\sigma_2}(p_1,p_2)\varphi^{*}_{\sigma'_1\sigma'_2}(p'_1,p'_2)
\end{equation}
Since we are only interested in the entanglement of the spin degrees of freedom we have to trace over the momenta of the two fermions, that is to say
\[\rho^{\text{free}}=\int d^3q_1\,d^3q_2\braket{q_1,q_2}{\Psi}\braket{\Psi}{q_1,q_2}\]
as $\braket{q_1,q_2}{p_1,p_2}=\bra0 b(q_1)b(q_2)b^\dagger(p_1)b^\dagger(p_2)\ket0=\delta(q_1-p_2)\delta(q_2-p_1)$\footnote{We are supposing distinguishable fermions to shorten notation} we obtain
\begin{equation}\label{rholibre}\rho_{\sigma_1\sigma_2,\sigma_1'\sigma_2'}^{\text{free}}=\int d^3p_1\,d^3p_2\,S_{\sigma_1\sigma_2,\sigma_1'\sigma_2'}(p_1,p_2,p_1,p_2)\end{equation}
If we want to calculate the degree of entanglement of this kind of states we could apply any of the common entanglement measures on the density matrix \eqref{rholibre}.

\section{Asymptotic Dynamics}

\subsection{Gauge Invariance and Asymptotic Evolution}

One formulation of the QED lagrangian which is very useful in order to discuss its gauge invariance is \cite{b6}
\[\mathcal{L}=\frac14 F^{\mu\nu}F_{\mu\nu}+ i\bar\psi\gamma^\mu\left(\partial_\mu-ieA_\mu\right)\psi -m\bar\psi\psi+\frac12\alpha B^2 +B\partial_\mu A^\mu\]
Where $B$ is a gauge fixing field \cite{b6}. The fields that appear in the lagrangian are (in the interaction picture)
\[\psi(x)=\frac{1}{(2\pi)^{\frac32}}\int d^3 p\sqrt{\frac{m}{p_0}}\sum_n \left[b_n(p)u_n(\vec p\,)e^{-ipx}+d^\dagger_n(p) v_n(\vec p\,)e^{ipx}\right]\]
\[\bar\psi(\vec x)=\frac{1}{(2\pi)^{\frac32}}\int d^3 p\sqrt{\frac{m}{p_0}}\sum_n \left[b_n^\dagger(p)\bar u_n(\vec p\,)e^{ipx}+d_n(p) \bar v_n(\vec p\,)e^{-ipx}\right]\]
\[A_\mu(x)=\frac{1}{(2\pi)^{\frac32}}\int \frac{d^3 k}{(2k_0)^{\frac12}}\left[a_\mu^\dagger(\vec k)e^{ikx}+a_\mu(\vec k) e^{-ikx}\right]\]
\[B(x)={(2\pi)^{\frac32}}\int \frac{d^3k}{\sqrt{2k_0}}\left[c^\dagger(k)e^{ik\cdot x}+c(k)e^{-ik\cdot x}\right]\]

The gauge fixing field allows us to identify physical states and observables:
\begin{itemize}
\item $c(k)\ket{\phi}=0\Leftrightarrow \ket{\phi}$ could be a physical state (since it is gauge invariant)
\item Let $O$ be an operator corresponding to an observable, then $\cnm{B}{O}=0$
\end{itemize}
It can be shown \cite{b4} that  in the Feynman Gauge ($\alpha=1\; ; \; B=-\partial_\mu A^\mu$) and for asymptotic times, the field modes of the $B$ field that annihilates physical states are given by the modes of the free vector potential:
\[c(k)=ik\cdot a(k)\]
that immediately leads us to the conclusion that the states $b_\sigma(p)\ket0$ are physical in the sense of gauge invariance.

However, given that the large $t$ limit of the QED Hamiltonian is not the free Hamiltonian, the states $b^\dagger\ket0$ do not correspond to the asymptotic states of QED. A well-defined physical asymptotic state must not only be gauge invariant but evolve with the proper Hamiltonian corresponding to the theory which describes its dynamics. We can conclude that the standard interaction picture does not serve to describe the correct asymptotic operators.

Thereby, the state $b^\dagger_\sigma(p)\ket{0}$, with $b^\dagger_\sigma(p)$ in the interaction picture, is not adequate since (although it is gauge invariant) it does not correspond to a state that evolves properly at asymptotic times.

Our goal is to find the charged particle creation and annihilation operators with well-defined momentum and spin that behave adequately at large $t$ (evolving undergoing an asymptotic interaction corresponding to asymptotic QED) and, furthermore, that they be gauge invariant (they must commute with $B$-field). To accomplish this objective we will need to compute the interaction operator in the large $t$ regime.

\subsection{The Asymptotic Interaction Operator in QED}

Let us show what we have announced before, that is to say, the  asymptotic Hamiltonian of QED is not $H_0$, to do this, we begin from the interaction operator between the charged fermions and the electromagnetic field
\begin{equation}\label{1}H^{\text{int}}=\int J^\mu(\vec x)A_\mu (\vec x) d^3 x=-e\int:\bar\psi(\vec x)\gamma^\mu\psi(\vec x):A_\mu(\vec x)d^3 x\end{equation}

After substituting the fields expresions in \eqref{1} and integrate over $x$, the resulting expression for $H^{\text{int}}(t)$ is an integral over the momenta  $\vec p,\vec q, \vec k$ of the fermions and the photons, whose are related by the conservation equation $\vec q=\vec p+\vec k$.

We divide the resulting terms after normal ordering expression \eqref{1} in two different groups:
\begin{itemize}
\item 1st group: terms that include two creators or two annihilators of charged particle $b^\dagger d^\dagger$ or $bd$, these terms are proportional to the following exponential factor
\[\exp(\pm ik_0t)\exp(ip_0t)\exp(iq_0t)=\exp\left[i\left(\sqrt{\vec p\,^2+m^2}+\sqrt{(\vec p+\vec k)^2+m^2}\pm k_0\right)t\right]\]
%o bien a
%\[\exp(\pm ik_0t)\exp(-ip_0t)\exp(-iq_0t)=\exp\left[-i\left(\sqrt{\vec p\,^2+m^2}+\sqrt{(\vec p+\vec k)^2+m^2}\mp k_0\right)t\right]\]
\item 2nd group: terms that include one creator and one annihilator of charged particle $b^\dagger b$ or $d^\dagger d$, these terms are proportional to the following exponential factor
\[\exp(\pm ik_0t)\exp(ip_0t)\exp(-iq_0t)=\exp\left[i\left(\sqrt{\vec p\,^2+m^2}-\sqrt{(\vec p+\vec k)^2+m^2}\pm k_0\right)t\right]\]
%o bien a
%\[\exp(\pm ik_0t)\exp(-ip_0t)\exp(iq_0t)=\exp\left[-i\left(\sqrt{\vec p\,^2+m^2}-\sqrt{(\vec p+\vec k)^2+m^2}\mp k_0\right)t\right]\]
\end{itemize}
The former acquire a phase that is highly oscillating at the limit $t\rightarrow\infty$, thus they cancel (in the weak sense) and can be neglected. The latter, on the other hand, have a non-zero contribución at the limit led by the value of the stationary phase, which is to say, at the proximity of $k=0$.

The leading contributions to the Interaction Hamiltonian in the asymptotic regime are
\begin{eqnarray*}H^{\text{int}}\!\!\!&=&\!\!\!\frac{-e}{(2\pi)^{\frac32}}\int m\sum_{n,m}\left[\bar u_m (\vec p\,)\gamma^\mu u_n(\vec p\,)b_m^\dagger (\vec p\,)b_n(\vec p\,)-\bar v_m (\vec q\,)\gamma^\mu v_n(\vec q\,)d_n^\dagger(\vec q\,)d_m(\vec q\,)\right]\cdot\\
&&\!\!\!\cdot\left[a_\mu^\dagger(-\vec k)+a_\mu(\vec k)\right]\delta(\vec p-\vec q + \vec k)e^{i(q_0-p_0-\omega)t}\frac{d^3pd^3qd^3k}{\sqrt{2q_0p_0\omega}}
\end{eqnarray*}
We will carry out the following approximation: let us consider $k\rightarrow0$ in all the functions that grow slowly enough with $t$, that is to say, the Dirac's Delta $\delta (\vec p-\vec q+\vec k)$ will fix $\vec p=\vec q$ for the spinors and the fermionic annihilation and creation operators, Thus we have that
\[i\hat u_n(\vec p+\vec k\,)\gamma^\mu u_{m}(\vec p\,)\rightarrow i\hat u_n(\vec p\,)\gamma^\mu u_{m}(\vec p\,)=\delta_{nm}\frac{p^\mu}{m}\qquad i\hat v_n(\vec p+\vec k\,)\gamma^\mu v_{m}(\vec p\,)\rightarrow i\hat v_n(\vec p\,)\gamma^\mu v_{m}(\vec p\,)=\delta_{nm}\frac{p^\mu}{m}\]
so that we have
\[H^{\text{int}}\!=\!\frac{-e}{(2\pi)^{\frac32}}\!\int \!\frac{d^3p\,d^3k}{\sqrt{2\omega}}\frac{p^\mu}{p_0}\! \sum_n\left[b_n(\vec p\,)b^\dagger_n(\vec p\,)+d_n(\vec p\,)d^\dagger_n(\vec p\,)\right]\!\left[a_\mu^\dagger(-\vec k)+a_\mu(\vec k)\right]e^{i\left(\sqrt{(\vec p+\vec k)^2+m^2}-\sqrt{\vec p\,^2+m^2}-\omega\right)\!t}\]
We could expand in power series of $k$ the exponential and keep the leading order (as the integral is dominated by $k\rightarrow0$)
\begin{eqnarray*}\sqrt{(\vec p+\vec k)^2+m^2}-\sqrt{\vec p\,^2+m^2}-\omega&\!\!\!=&\!\!\!\left.\sqrt{(\vec p+\vec k)^2+m^2}\right|_0+\vec k\cdot\nabla_{\vec k}\left.\sqrt{(\vec p+\vec k)^2+m^2}\right|_0 -\\
&\!\!\!&-\sqrt{\vec p\,^2+m^2} -\omega+O(\omega^2)=\frac{\vec k\!\cdot\!\vec p}{p_0}-\omega+O(\omega^2)\end{eqnarray*}
So that we can write the asymptotic interaction Hamiltonian the following way:
\begin{equation}\label{5}H^{\text{int}}_{\text{as}}(t)=-e\int \frac{d^3 k}{(2k_0)^{\frac12}} \int \frac{1}{(2\pi)^{\frac{3}{2}}}\frac{d^3 p}{p_0}p^\mu e^{i\frac{\vec p\cdot \vec k}{p_{_0}}t-i\omega t}\rho(\vec p\,)\left[a_\mu^\dagger(-\vec k)+a_\mu(\vec k)\right]\end{equation}
Which can be also written as
\begin{equation}\label{6a}H^{\text{int}}_{\text{as}}(t)=-e\int d^3 x\int d^3 k\int \frac{1}{(2\pi)^3}\frac{d^3 p}{p_0}p^\mu e^{i\frac{\vec p\cdot \vec k}{p_{_0}}t-i\omega t}e^{-ik\cdot x}\rho(\vec p\,)  a_\mu(x)\end{equation}
Where we have used that  $a_\mu(x)=a_\mu^{(+)} (x)+a_\mu^{(-)} (x)$ with \cite{b7}
\[a_\mu(k)=\frac{1}{(2\pi)^{3/2}}\int d^3 x\,\sqrt{2k_0}\, e^{-i k\cdot x} a_\mu^{(-)} (x)\qquad a_\mu^\dagger(k)=\frac{1}{(2\pi)^{3/2}}\int d^3 x\,\sqrt{2k_0}\, e^{i k\cdot x} a_\mu^{(+)} (x)\]
The integral over  $d^3k$ yields to a Delta, resulting that
\begin{equation}\label{6b}H^{\text{int}}_{\text{as}}(t)=-e\int J^\mu_{\text{as}}(\vec x,t)\,a_\mu(x)\,d^3 x\end{equation}
where
\begin{equation}\label{corriente}J^{\text{as}}_\mu(t,x)=\int d^3p\frac{p_\mu}{p_0}\rho(p)\delta^3\left(\vec x-\frac{\vec p}{p_0}t\right)\end{equation}
being
\[\rho(\vec p\,)=\sum_n \left[b_n^\dagger (\vec p\,) b_n(\vec p\,)-d_n^\dagger(\vec p\,)d_n(\vec p\,)\right]=\rho_-(\vec p)-\rho_+(\vec p)\]

In order to clarify future calculations, is convenient to write the interaction Hamiltonian \eqref{5} in the next form
\begin{equation}\label{Hutil}H^{\text{int}}_{\text{as}}(t)=-e\int \frac{d^3 k}{(2k_0)^{\frac12}} \frac{1}{(2\pi)^{\frac{3}{2}}}\hat J^\mu_{\text{as}}\left[a_\mu^\dagger(-\vec k)+a_\mu(\vec k)\right]\end{equation}
Where
\begin{equation}\label{Jutil}\hat J^\mu_{\text{as}}=\int d^3p\, \frac{p^\mu}{p_0} e^{-i\frac{p\cdot k}{p_{_0}}t}\rho(\vec p\,)\end{equation}

Summarizing : The Interaction operator is non-zero in the asymptotic regime. The residual interaction is equivalent to the interaction between a spinless charged particle current and the electromagnetic field. The resulting expression is universal, and can be readily generalized considering charges with an arbitrary spin.

Eventually, the operator $\hat J^\mu_{\text{as}}$ has a very clear physical meaning: a given state of (free) charged particles with a well-defined momentum
\begin{equation}\label{6}\ket{\Psi(\vec p_1s_1,\dots,\vec p_ns_n,\vec q_1i_1,\dots,\vec q_ni_n)}=b^\dagger_{s_1}(\vec p_1\,)\dots b^\dagger_{s_n}(\vec p_n)d^\dagger_{i_1}(\vec q_1\,)\dots d^\dagger_{i_n}(\vec q_n)\ket0\end{equation}
is an eigenstate of this operator, being the corresponding eigenvalue
\[\lambda = \sum_{j=1}^m j_\mu( k,t;q_j)+\sum_{j=1}^n j_\mu(k,t;p_j) \qquad\text{where}\qquad j_\mu(\vec k,t;p)=e\frac{p_\mu}{p_0}e^{-i\frac{k\cdot p}{p_0}t}\]
Furthermore, it is important tho emphasize that current operators in different points of the space-time commute: $\cnm{J^\mu_{\text{as}}(x)}{J^\mu_{\text{as}}(y)}=0$.

\subsection{The Asymptotic Interaction Picture}\label{sechei}

Perturbation Theory is based on the fact of consider that, at asymptotic times, the interaction is deactivated and, therefore, $H(t\rightarrow\infty)=H_0$ thus, the ``in'' y ``out'' states that would evolve by means of the free Hamiltonian. This fact allow us to include this asymptotic temporal dependence within the operators, building the so-called standard interaction picture in terms of the Heisenberg picture:
\[O_I(t,t_0)=\operatorname{T}\exp\left(-i\int_{t_0}^t d\tau \left[H(\tau)-H_{0}(\tau)\right]\right)O_{H}(t)\operatorname{\tilde T}\exp\left(i\int_{t_0}^t d\tau \left[H(\tau)-H_{0}(\tau)\right]\right).\]
This transformation requires the introduction of an arbitrary time $t_0$ in which the fields in both pictures are equal: $O_H(t_0)=O_I(t_0,t_0)$. This time $t_0$ is completely arbitrary since if we transform back our operator into the Heisenberg picture, every explicit reference to $t_0$ will vanish.

However, we have seen that, in QED, the asymptotic limit of the Hamiltonian is not the free Hamiltonian but $H^{\text{as}}=H_{\text{int}}^{\text{as}}+H_0$, Thus, it is necessary to build a new interaction picture in which the operators include the proper temporal dependence at asymptotic times. We build it as follows
\[O^{\text{as}}_I(t,t_0)=\operatorname{T}\exp\left(-i\int_{t_0}^t d\tau \left[H(\tau)-H^{as}(\tau)\right]\right)O_{H}(t)\operatorname{\tilde T}\exp\left(i\int_{t_0}^t d\tau \left[H(\tau)-H^{as}(\tau)\right]\right)\]
Or, otherwise
\[O^{\text{as}}_I(t,t_0)=\operatorname{T}\exp\left(-i\int_{t_0}^t d\tau \left[H_{\text{int}}(\tau)-H_{\text{int}}^{as}(\tau)\right]\right)O_{H}(t)\operatorname{\tilde T}\exp\left(i\int_{t_0}^t d\tau \left[H_{\text{int}}(\tau)-H_{\text{int}}^{as}(\tau)\right]\right)\]
In order to study the dependence on $t_0$ of the previous expression, let us expand the exponentials to first order
\[O^{\text{as}}_I(t,t_0)=O_H(t)+i\int_{t_0}^td\tau\,\cnm{H^{\text{as}}_{\text{int}}(\tau)-H_{\text{int}}(\tau)}{O_H(t)}\]
Here, we clearly see that there is an explicit dependence on $t_0$ that comes from the lower integration limit. This dependence is governed by the difference
\[H^{\text{as}}_{\text{int}}(t_0)-H_{\text{int}}(t_0)\]
So if we choose one $t_0$ that be itself an asymptotic time (large enough) the dependence on $t_0$ will vanish by construction. Thus, we can set $t_0=\pm\infty$. The analogous can be demonstrated for every order in the expansion.

\subsection{Building the Asymptotic Fields}

In this section, we will go from the standard interaction picture to the new asymptotic interaction picture (via the usual time ordered exponential of the difference between the complete Hamiltonian and the asymptotic one) Thus, the field operators will include the asymptotic evolution. Given the functional form of the asymptotic Hamiltonian, the fact that the asymptotic current operators commute and that $[a^\dagger, a]$ is a C-number, it can be proved that
\begin{equation}\label{2}\operatorname{T}\exp\left(i\int_{t_0}^t d\tau H^{\text{as}}_{\text{int}}(\tau)\right)=\exp\left(i\int_{t_0}^t d\tau H^{\text{as}}_{\text{int}}(\tau)\right)\exp\left(\frac12 \int_{t_0}^{t} d\tau_1 \int_{t_0}^{\tau_1}d\tau_2\cnm{ H^{\text{as}}_{\text{int}}(\tau_1) }{ H^{\text{as}}_{\text{int}}(\tau_2)}\right)\end{equation}
We can now obtain the field operators (fermionic and photonic) in the asymptotic interaction picture.

Concerning the photon field, provided that
$\cnm{\cnm{H_{\text{int}}^{\text{as}}(\tau_1)}{H_{\text{int}}^{\text{as}}(\tau_2)}}{A_\mu(x)}=0$, is easy to build the asymptotic vector potential  operator,
\[A_\mu^{\text{as}}(x)=\exp\left(i\int_{t_0}^td\tau H^{\text{as}}_{\text{int}}(\tau)\right)A_\mu(x)\exp\left(-i\int_{t_0}^td\tau H^{\text{as}}_{\text{int}}(\tau)\right)\]
so we have that
\[A^{\text{as}}_\mu(x)=A_\mu(x)-e\int_{t_0}^t d\tau  d^3y D(\tau-t,\vec y-\vec x) J^{\text{as}}_\mu (\tau,\vec y)\]
It is important to remark that, from this expression, we obtain $\square A^{\text{as}}_\mu(x)=-eJ^{\text{as}}_\mu(x)$, that is to say, this asymptotic vector potential contains the Coulomb field generated by the asymptotic charged current $J^{\text{as}}_\mu$ plus the free field. This is reasonable since at asymptotic times the residual interaction is like the interaction between a charged current and the electromagnetic field.

Furthermore, given that $J^\mu(x)$ commute with itself, the asymptotic bosonic field obeys the same commutator as the free field
\[\cnm{A^{\text{as}}_\mu(x)}{A^{\text{as}}_\nu(y)}=-ig_{\mu\nu}D(x-y)\]

Let us consider the matter field, in this case both terms in \eqref{2} contribute. The computations are not difficult but a little bit tiresome (above all for the phase factor), the detail can be found in \cite{b2} and \cite{b7}. The result is that there exists a transformation $U(t)$ that goes from the standard interaction picture to the asymptotic interaction picture which is given by
\[U(t)=\exp\left[R(t)\right]\exp\left[i\Phi(t)\right]\]
Where we will call the operators $\exp\left[R(t)\right]$ and $\exp\left[i\Phi(t)\right]$ distortion and phase operators respectively\footnote{ Notice that we have set $t_0\rightarrow\pm\infty$ through the calculations due to the arguments argued in section \ref{sechei} (as it is directly done in  \cite{b2}, \cite{b4} and \cite{b7}) except for the phase operator, in which we will keep the explicit dependence on $t_0$ because its contribution is apparently divergent for asymptotic values of $t_0$. We will show that the dependence on $t_0$ in the phase operator is irrelevant, and the dressing procedure will cancel it.} which are given by
\begin{equation}\label{3}R(t)=\frac{e}{(2\pi)^{\frac{3}{2}}}\int \frac{p^\mu}{pk}\left[a_\mu^\dagger(\vec k)e^{i\frac{k\cdot p}{p_0}t}-a_\mu(\vec k)e^{-i\frac{k\cdot p}{p_0}t}\right]\rho(\vec p\,)d\vec p\frac{d\vec k}{(2k_0)^{\frac12}}\end{equation}
\begin{equation}\label{4}\Phi(t)=\frac{e^2}{8\pi}\int :\rho(\vec p\,)\rho(\vec q\,):\frac{p\!\cdot\!q}{\sqrt{(pq)^2-m^4}}\operatorname{sign}(t)\ln\frac{|t|}{t_0} d\vec p d\vec q\end{equation}

\subsection{Asymptotic Evolution of the two Particles State}\label{secevol}

For our purposes, we would like to define a two charged particles state (eigenstate of the fermionic number operator with eigenvalue 2) with well-defined momenta and spin projection. In the standard interaction picture such a state would have the form
\[\ket\Psi=b_{\sigma_2}^\dagger(p_2)b_{\sigma_1}^\dagger(p_1)\ket0\]
where the operators include the free evolution within.

Nonetheless, as we argued above, the interaction picture is not adequate to describe the dynamical behaviour of charged particles in the large t regime. If we transform the operators to the asymptotic interaction picture we will have
\[\ket{\Psi}={b^{\text{as}}_{\sigma_2}}^\dagger(p_2){b^{\text{as}}_{\sigma_1}}^\dagger(p_1)\ket0=\exp\left[R(t)\right]\exp\left[i\Phi(t)\right]b_{\sigma_2}^\dagger(p_2)b_{\sigma_1}^\dagger(p_1)\ket0\]
Taking into account that the fermionic Fock states are eigenstates of the asymptotic current operator, It can be obtained an expression in which the asymptotic factors are free from fermionic operators
\begin{equation}\label{estadoas}
\ket\psi_{\text{as}}=e^{i\phi(u_r,t)}e^{W(p_1,p_2,t)}b^\dagger_{\sigma_{_1}}(p_1)b^\dagger_{\sigma_{_2}}(p_2)\ket0
\end{equation}
Where
\begin{equation}\label{vdoble}W(p_1,p_2,t)=\frac{e}{(2\pi)^{\frac32}}\int\frac{d^3k}{\sqrt{2k_0}}\left[\left(\frac{p_1^\mu}{p_1 k}e^{i\frac{k p_1}{{p_1}_{_0}}t}+\frac{p_2^\mu}{p_2 k}e^{i\frac{k p_2}{{p_2}_{_0}}t}\right)a_\mu^\dagger(k)-\left(\frac{p_1^\mu}{p_1 k}e^{-i\frac{k p_1}{{p_1}_{_0}}t}+\frac{p_2^\mu}{p_2 k}e^{-i\frac{k p_2}{{p_2}_{_0}}t}\right)a_\mu(k)\right]\end{equation}
\begin{equation}\label{divfase}\phi=\frac{e^2}{4\pi}u^{-1}_r(p_1,p_2)\log\frac{|t|}{t_0}\end{equation}
are the ``eigenvalues'' of the $R(t)$ y $\Phi(t)$ operators  associated to the considered state \cite{b2} and where  $u_r(p_1,p_2)$ is the absolute value of the relative velocity between the two particles, $u(p,q)=\sqrt{1-m^4/(pq)^2}$.

It is obvious that this construction is not Gauge invariant since $W(p_1,p_2,t)$ contains exponentials of photon operators. This fact could make us think that the charged particle concept could not be well-defined.

Furthermore, the resulting phase  factor \eqref{divfase} (that is not often considered in the literature as it is of second order and it does not appear in the cross sections) Although it is gauge invariant, it seems to be divergent for $t_0\rightarrow\pm\infty$. In the following sections we will demonstrate that this apparent divergence disappears when we apply the ``dressing'' procedure building physical gauge invariant states.

To end this section let us mention an interesting result that we will use later: It can be proved \cite{b2} that the transformation of the asymptotic state $\ket{\psi_{\text{as}}}\rightarrow\ket{\psi'_{\text{as}}}$ such that
\begin{equation}\label{invasin}\ket{\psi'_{\text{as}}}=e^{\phi(u_r,t)}e^{W'(p_1,p_2,c_1,c_2,t)}b^\dagger_{\sigma_{_1}}(p_1)b^\dagger_{\sigma_{_2}}(p_2)\ket0\end{equation}
where
\begin{eqnarray*}
W'(p_1,p_2,c_1,c_2,t)&\!\!\!=&\!\!\!\frac{e}{(2\pi)^{\frac32}}\int\frac{d^3k}{\sqrt{2k_0}}\left[\left(\left\{\frac{p_1^\mu}{p_1 k}-c_1^\mu\right\}e^{i\frac{k p_1}{{p_1}_{_0}}t}+\left\{\frac{p_2^\mu}{p_2 k}-c_2^\mu\right\}e^{i\frac{k p_2}{{p_2}_{_0}}t}\right)a_\mu^\dagger(k)-\right.\\
&-&\!\!\!\
\left. \left(\left\{\frac{p_1^\mu}{p_1 k}-c_1^\mu\right\}e^{-i\frac{k p_1}{{p_1}_{_0}}t}+\left\{\frac{p_2^\mu}{p_2 k}-c_2^\mu\right\}e^{-i\frac{k p_2}{{p_2}_{_0}}t}\right)a_\mu(k)\right]
\end{eqnarray*}
(that differentiates from $W$ in the terms $c_1^\mu$ and $c_2^\mu$) operates within the asymptotic states Hilbert $\mathcal{H}_{\text{as}}$, that is to say
\[\ket{\psi_{\text{as}}}\in\mathcal{H}_{\text{as}}\Leftrightarrow \ket{\psi'_{\text{as}}}\in\mathcal{H}_{\text{as}}\]
If the following integrals converge
\begin{equation}\label{cond1}\int\left[c^i_\mu(\vec p,\vec k)-\frac{p_\mu}{pk}e^{i\frac{kp}{p_{_0}}t}\right]\left[f_\mu^*(\vec p,\vec k)-\frac{p_\mu}{pk}e^{-i\frac{kp}{p_{_0}}t}\right]\frac{d^3k}{2k_0}\qquad \text{and}\qquad\int\left|c_\mu^{i*}(\vec q,\vec k)\frac{q_\mu}{pk}e^{i\frac{kq}{q_{_0}}t}-f_\mu(\vec q,\vec k)\frac{q_\mu}{qk}e^{-i\frac{kq}{q_{_0}}t}\right|\frac{d^3k}{2k_0}.\end{equation}
%\begin{equation}\label{cond2}\int\left|c_\mu^{i*}(\vec q,\vec k)\frac{q_\mu}{pk}e^{i\frac{kq}{q_{_0}}t}-f_\mu(\vec q,\vec k)\frac{q_\mu}{qk}e^{-i\frac{kq}{q_{_0}}t}\right|\frac{d^3k}{2k_0}.\end{equation}
This is a formal result which we will use in this work to understand certain aspects of the ``dressing'' procedure.

\section{Building Physical Asymptotic States}

\subsection{The ``dressing'' procedure}\label{Sdressing}

The states built so far (that evolve correctly in the asymptotic regime) are not physical (they are not gauge invariant). The solution to recover the physical particle conception \cite{b4},\cite{b5} consists on ``dressing'' the fields by means of operators that restore the gauge invariance without altering their dynamics. This procedure that allows us to define a physical particle state in the asymptotic regime.

The idea is to ``dress'' the matter field with a ``dressing'' operator that under gauge transformations behaves opposite to the field and hence the built object will be gauge invariant. This is to say that given a field $\varphi(x)$ that under gauge transformations behaves as follows
\[\varphi(x)\longrightarrow \varphi(x) e^{ie\theta(x)}\]
we need a dressing such that
\begin{equation}\label{6a}h^{-1}(x)\longrightarrow h^{-1}(x) e^{-ie\theta(x)}\end{equation}
so that the dressed field
\begin{equation}\label{6b}\Phi(x)=h^{-1}(x)\varphi(x)\end{equation}
is gauge invariant.

But this is not the only condition that the dressing should satisfy: We will also demand that it conserves the dynamics of the field that we are dressing. First of all let us consider a free matter field in the infinite-mass limit it can be proved \cite{b8} that velocity is superselected and the equation of motion for the field has the universal form
\begin{equation}\label{7}u\cdot\partial\Phi(x)=0\end{equation}
where $u$ is the four-velocity of the heavy particle. This equation is simply the statement that the field is constant along the world line of a particle movin with 4-velocity $u$. Consequently, if we parametrise the world line of a particle that moves with a 4-velocity $u^\mu=\gamma(\eta+v)$ where
\begin{itemize}
\item $\eta$ is an unitary temporal vector $\eta=(1,\vec 0)$
\item $v=(0,\vec v)$ being $\vec v$ the 3-velocity of the particle
\item $\gamma=\left(1-|\vec v|^2\right)^{-1/2}$
\end{itemize}
we obtain that
\begin{equation}\label{pripar}x^\mu(s)=x^\mu+(s-x^0)(\eta+v)^\mu\end{equation}
Then, \eqref{7} implies that, for an arbitrary $s$, $\Phi\left[x(s)\right]=\Phi\left[x(0)\right]$.

If the matter field is minimally coupled to the EM field the equation of motion turns into
\begin{equation}\label{8}u\cdot D\Phi(x)=0\qquad\text{where } D_\mu=\partial_\mu-ieA_\mu.\end{equation}

This analysis, which is done for the heavy matter sector, is also valid for any field that asymptotically could correspond to a charged particle with four-velocity $u$. We will give two arguments to justify this statement:
\begin{enumerate}
\item The asymptotic dynamics of QED is governed by soft photons for whom any electron is heavy.
\item It can be shown \cite{b9} that the asymptotic interaction Hamiltonian vanishes for the propagator of the dressed fields that satisfy  \eqref{8} if one is at the correct point in the mass-shell.
\end{enumerate}
If we demand the dressing preserves the heavy particle dynamics (we impose \eqref{8} to \eqref{6b}) we will obtain the next equation for the dressing operator \cite{b4} (that is called ``the dressing equation''):
\begin{equation}\label{9}u\cdot\partial h^{-1}(x)=-ie\, h^{-1}(x)\, u\cdot A(x)\end{equation}
Summarizing, the equations \eqref{6a} and \eqref{9} are the two fundamental requirements that the dressed field $\Phi$ must satisfy in order to describe physical charged particles.

\subsection{``Dressing'' the two particles state}

The dressing operator $h^{-1}$ consists of  a phase term and a distortion term (so the asymptotic evolution operator does) \cite{b4} that is to say, we can express
\[h^{-1}(x)=e^{\chi_i(x)}e^{-iK_i(x)}\]
where it can be shown \cite{b4} that in the large $t$ limit
\begin{equation}\label{dressingxi}\chi_i(x)=\int\frac{d^3k}{(2\pi)^3}\frac{1}{\sqrt{2k_0}}\left(\frac{V^\mu a_\mu(k)}{V\!\cdot\! k}e^{-ik\cdot x}-\frac{V^\mu a^\dagger_\mu(k)}{V\!\cdot\! k}e^{ik\cdot x}\right)\end{equation}
where
\[V_i^\mu=(\eta+v_i)^\mu(\eta-v_i)\cdot k-k^\mu\]
And also that
\begin{equation}\label{dressingphase}e^{-iK_i}=\exp\left\{-ie\int_{t_0}^{t} (\eta+ v_i)^\mu\frac{\partial^\nu F_{\nu\mu}}{\mathcal{G}\cdot\partial}\left[x(s)\right]ds\right\}\end{equation}
where we have an integral along the world line of the massive particle with constant 4-velocity $u_i^\mu$ parametrised by means of $s$ \cite{b4}

The action of the operator $\frac{1}{\mathcal{G}\cdot\partial}$ is defined \cite{b4}, \cite{b5} as
\[\frac{1}{\mathcal{G}\cdot\partial}f(\vec x)\equiv \int d^3 z G(\vec x-\vec z)f(\vec z)\!\!\qquad \text{where}\!\!\qquad G(x)=-\int\frac{d^3k}{(2\pi)^3}\frac{1}{\vec k^2-(\vec v \cdot \vec k)^2}e^{i\vec k\cdot\vec x}=-\frac{1}{4\pi}\frac{\gamma}{\sqrt{\vec x\,^2+\gamma^2 (\vec v \cdot \vec x)^2}}.\]

Given that the dressed matter field has the expression $\Phi(x)=h^{-1}(x)\varphi(x)$, the dressed creation operators in function of the dressed asymptotic field is given by
\[b^\dagger_{\text{as,d}}(p,\sigma)=\int d^3x \sqrt{\frac{m}{p_0}} u_\sigma^{\dagger}(p)\underbrace{e^{-i  \hat K_i(x)}e^{  \hat \chi(x)}\varphi^{\text{as}}(x)}_{  \hat \Phi^{\text{as}}(x)}e^{iq\cdot x}\]
that is to say
\[b_\text{as,d}^\dagger(q,\sigma)=(2\pi)^{-3}\sum_{\sigma'}\int d^3q'\sqrt{\frac{m^4}{E_q'E_q}}u_\sigma(q)u_{\sigma'}(q')\!\int d^3x\,e^{i(q-q') x} e^{\chi_i(x)}e^{-iK_i(x)}b^\dagger_\text{as}(q',\sigma')\]
Expressing the asymptotic creation operator in terms of the creation operator in the standard interaction picture:
\[b_\text{as,d}^\dagger(q,\sigma)=(2\pi)^{-3}\sum_{\sigma'}\int d^3q'\sqrt{\frac{m^4}{E_q'E_q}}u_\sigma(q)u_{\sigma'}(q')\!\int d^3x\, e^{i(q-q') x} e^{\chi_i(x)}e^{-iK_i(x)}\underbrace{e^{R(t)}e^{i\Phi}b^\dagger(p')e^{-i\Phi}e^{-R(t)}}_{b_{\text{as}}^\dagger (q')}\]
It can be demonstrated that $e^{-iK_i(x)}$ commutes with $e^{R(t)}e^{i\phi(t)}$ (since current operators commute) so that
\[b_\text{as,d}^\dagger(q,\sigma)=(2\pi)^{-3}\sum_{\sigma'}\int d^3q'\sqrt{\frac{m^4}{E_q'E_q}}u_\sigma(q)u_{\sigma'}(q')\!\int d^3x\, e^{i(q-q') x} e^{\chi_i(x)}e^{R(t)}e^{i\Phi(t)}e^{-iK_i(x)}b^\dagger(q',\sigma')e^{-i\Phi}e^{-R(t)}\]

Let us start computing the phase operator:
The relevant dynamics of the gauge fields in the asymptotic regime is given by \cite{b12},\cite{b6}:
\[\partial^\nu F^{\text{as}}_{\nu\mu}=\partial_\mu B - eJ^{\text{as}}_\mu\]
where we see that the matter coupling comes from the asymptotic current. Using this and omitting the dependences with the gauge fixing field, that are irrelevant acting on physical states, estados físicos, we can write $K_i$ as
\[K_i=\frac{e^2}{4\pi}\int_{t_0}^{t}ds (\eta+ v_i)^\mu\int d^3 z \frac{\gamma\,J^{\text{as}}_\mu(\vec z, s)}{\sqrt{\{\vec x(s)-\vec z\}^2+\gamma^2 [\vec v \cdot  \{\vec x(s)-\vec z\}]^2}} \]
Substituting the asymptotic current \eqref{corriente} we obtain that
%\[J^{\text{as}}_\mu(t,x)=\int d^3p\frac{p_\mu}{p_0}\rho(p)\delta^3\left(\vec x- \vec x(0)-\frac{\vec p}{p_0}t\right)\]
%sustituimos en $K_i$ y operando %hacemos la integral en $d^3z$ con la delta
\begin{equation}\label{Ki}
K_i=\frac{e^2}{4\pi}\int_{t_0}^{t}ds (\eta+ v_i)^\mu\int d^3p\, \gamma\frac{p_\mu}{p_0}\rho(p)f(s,\vec v_i,\vec x_i)
\end{equation}
where
\begin{equation}\label{efe}f(s,\vec v_i,\vec x_i,p_j)=\left\{\left[\vec x(s) - s\frac{\vec p_j}{p_{j0}}\right]^2+\gamma^2 \left[\vec v_i \cdot  \left( \vec x(s) + s\frac{\vec p_j}{p_{j0}}\right)\right]^2\right\}^{\frac{-1}{2}}\end{equation}
We need to commute $e^{-iK_i(x)}$ with $b^\dagger$, in order to do it, we will use Hadamard's Lemma:
\[e^{xA}Be^{-xA}=B+\cnm{A}{B}x+\frac{1}{2!}\cnm{A}{\cnm{A}{B}}x^2+\dots\]
Since we have that
\[e^{-iK_i}b^\dagger(q')=e^{-iK_i}b^\dagger(q')e^{iK_i}e^{-iK_i}\]
Let us use Hadamard's Lemma with
\[A=K_i\qquad B=b^\dagger(q')\qquad x=-i\]
\[e^{-iK_i}b^\dagger(q')e^{iK_i}=\left(b^\dagger(q')-i\cnm{K_i}{b^\dagger(q')}+(-i)^2\frac{1}{2!}\cnm{K_i}{\cnm{K_i}{b^\dagger(q')}}+\dots\right)\]
And the commutator is
\[\cnm{K_i}{b^\dagger(q')}=\frac{e^2}{4\pi}\int_{t_0}^{t} (\eta+ v_i)^\mu\int d^3p\, \gamma\frac{p_\mu}{p_0}\cnm{\rho(p)}{b^\dagger(q')}f(s,\vec v_i,\vec x_1) ds\]
Since we know that
\[\cnm{\rho(p)}{b^\dagger(q')}=\cnm{b^\dagger(p)b(p)}{b^\dagger(q')}=b^\dagger(q')\delta(\vec p-\vec q')\]
we can perform the integration over $d^3p$, resulting
\begin{equation}\label{kappadef}\cnm{K_i}{b^\dagger(q')}=\frac{e^2}{4\pi}\int_{t_0}^{t} (\eta+ v_i)^\mu \gamma\frac{{q'}_{\mu}}{{q'}_{_0}}b^\dagger(q')f(s,\vec v_1,\vec x_1) ds\equiv \kappa(v_i,q', t) b^\dagger(q')\end{equation}
On the mass shell of $q'$, it is also satisfied that $\vec v_i=\vec q'/q_0'$, so that, (given that $q^\mu/q_0=(\eta+v)^\mu$ and $(\eta+ v_i)^\mu (\eta+ v_i)_\mu =\frac{1}{\gamma^2}$)
\[\kappa_i=\frac{e^2}{4\pi}\frac{1}{\gamma}\int_{t_0}^{t}ds f(s,\vec v_i',\vec x_i,q')\]
Therefore, the dressed asymptotic creation operator has the form
\[b_\text{as,d}^\dagger(q,\sigma)=(2\pi)^{-3}\sum_{\sigma'}\int d^3q'\sqrt{\frac{m^4}{E_q'E_q}}u_\sigma(q)u_{\sigma'}(q')\!\int d^3x\, e^{i(q-q') x} e^{\chi_i(x)}e^{R(t)}e^{i\Phi(t)}e^{-i\kappa_i(x)}b^\dagger(q',\sigma')e^{-iK_i(x)}e^{-R(t)}e^{-\Phi(t)}\]
For our purposes we need to calculate the product of two dressed asymptotic creation operators applied to the vacuum state
\begin{equation*}
\begin{split}
b_\text{as,d}^\dagger(q_1,\sigma_1)b_\text{as,d}^\dagger(q_1,\sigma_1)\ket0 &  =(2\pi)^{-6}\sum_{\sigma\sigma'}\int d^3q_1'd^3q_2'\, \sqrt{\frac{m^4}{E'_{q_1} E_{p_1'}E'_{q_2} E_{p'_2}}}u^\dagger_{\sigma}(q)u_{\sigma'}(q')u^\dagger_{\sigma}(q)u_{\sigma'}(q')\,\cdot \\
& \cdot \int d^3x_1\,d^3x_2\, e^{i(q_1-q_1')x_1}e^{i(q_2-q_2')x_2}e^{\chi_1(x)}e^{\chi_2(x)}e^{R(t)}e^{i\Phi}e^{-i\kappa_1(x)}e^{-i\kappa_2(x)}b^\dagger(q_1')e^{-iK_{1}(x)} b^\dagger(q_2')\ket0
\end{split}
\end{equation*}
To arrive to this expression we have used that $e^{-iK_i}\ket0=\ket0$, $i=1,2$ and we have performed two commutations\footnote{We have commuted $e^{-iK_1}$ with $b^\dagger(q'_1)$ and $e^{-iK_2}$ with $b^\dagger(q'_2)$, giving us the two C-numbers $e^{-i\kappa_1}$ and $e^{-i\kappa_2}$}.  Now, we need to commute $e^{-iK_{1}(x)}$ with $b^\dagger(q_2')$, the computation is analogous to the other two performed commutations with the only difference that the velocity that appears in the operator is on the mass shell of $q'_1$ and we are commuting with the creator with momentum $q'_2$, the calculations yields to the term
\[\kappa_{12}=\frac{e^2}{4\pi}\gamma\int_{t_0}^{t}ds\, (\eta+ v_1)^\mu (\eta+ v_2)_\mu \gamma f(s,\vec v_1,\vec x_1,q'_2) =\frac{e^2}{4\pi}\gamma(1-\vec v_1\!\cdot\!\vec v_2) \int_{t_0}^{t} ds\, f(s,\vec v_1,\vec x_1,q'_2)\]
Therefore, the two charged fermions state has the form
\begin{equation*}
\begin{split}
b_\text{as,d}^\dagger(q_1,\sigma_1)b_\text{as,d}^\dagger(q_1,\sigma_1)\ket0 &  =(2\pi)^{-6}\sum_{\sigma\sigma'}\int d^3q_1'd^3q_2'\, \sqrt{\frac{m^4}{E'_{q_1} E_{p_1'}E'_{q_2} E_{p'_2}}}u^\dagger_{\sigma}(q)u_{\sigma'}(q')u^\dagger_{\sigma}(q)u_{\sigma'}(q')\,\cdot \\
& \cdot \int d^3x_1\,d^3x_2\, e^{i(q_1-q_1')x_1}e^{i(q_2-q_2')x_2}e^{\chi_1(x)}e^{\chi_2(x)}e^{R(t)}e^{i\Phi}e^{-i\kappa_1(x)}e^{-i\kappa_2(x)}e^{-i\kappa_{12}(x)}b^\dagger(q_1') b^\dagger(q_2')\ket0
\end{split}
\end{equation*}

The next step is to compute the integrals over the affine parameter $s$ along the world line of the massive particles, in order to compute the phase factors we will parametrise it in the following way
\[x^\mu(s)=x^\mu+(s-t)(\eta+v)^\mu\Rightarrow\left\{\begin{array}{l}
\vec x(s) = \vec x+ (s-t)\vec v\\
x^0(s)=s
\end{array}\right.\]
\[\kappa_1=\frac{e^2}{4\pi}\frac{1}{\gamma}\int_{t_0}^{t}ds \left\{\left(\vec x -\vec v' t \right)^2+\gamma^2 \left[\vec v' \cdot  \left( \vec x-\vec v' t \right)\right]^2\right\}^{\frac{-1}{2}}\]
integrating trivially
\[\kappa_1=\frac{e^2}{4\pi}\frac{1}{\gamma}\frac{t-t_0}{\sqrt{\left(\vec x -\vec v' t \right)^2+\gamma^2 \left[\vec v' \cdot  \left( \vec x-\vec v' t \right)\right]^2}}\]
The term in the denominator is a retarded position $\vec R=\vec x-\vec vt$ analogous to the retarded position in the Lienard-Wiechert potentials \cite{b13} (we will discuss this point later):
\begin{equation}\label{kappa1}\kappa_1=\frac{e^2}{4\pi}\frac{1}{\gamma}\frac{t-t_0}{\sqrt{R_1^2+\gamma^2 \left[\vec v' \cdot  \left( R_1 \right)\right]^2}}\end{equation}

At large times, the leading contribution to $R$ is the term $-vt$ since we are considering the region of the space-time in which we have large times but finite positions as in the spatial infinite the interaction actually vanishes (see discussion in \cite{b4}). Taking it into account, the contribution of $x$ is subdominant and we can assume that we are always at positions given by the asymptotic times so that the phase can be factored out from the integral over $x$ in the asymptotic regime approximation.

Analogously, we obtain that $\kappa_2$ is
\begin{equation}\label{kappa2}\kappa_2=\frac{e^2}{4\pi}\frac{1}{\gamma}\frac{t-t_0}{\sqrt{\vec R_2^2+\gamma^2 \left[\vec v' \cdot  \left( \vec R_2 \right)\right]^2}}\end{equation}

Let us show that the phase factors \eqref{kappa1}, \eqref{kappa2} have a clear physical interpretation, writing the Liénard-Wiechert potential created by a moving charge with 4-velocity $u^\mu=(\gamma,\vec u)$ at large distances (cf. \cite{b13} Cap. 14):
\[A^{\text{class}}_\mu=-\frac{e}{4\pi}\frac{u_\mu}{\sqrt{R^2+\left[\vec u\!\cdot\!\vec R\right]^2}}=-\frac{e}{4\pi}\frac{\gamma (\eta+v)_\mu}{\sqrt{R^2+\gamma^2\left[\vec v\!\cdot\!\vec R\right]^2}}\]
Where $R$ is the retarded position. On the other hand, the asymptotic current is
\[J^{\text{as}}_\mu(t,x)=\int d^3p\,(n+v)_\mu\rho(p)\delta^3\left(\vec x -\vec v t\right)\]
acting with it on a state $b^\dagger(p)\ket0$ we obtain the next eigenvalue:
\[j^{\text{as}}_\mu(t,x)=(n+v)_\mu \delta^3\left(\vec x-\vec v\,t\right)\]
%Sustituyendo $x$ por la parametrización
%\[x(t)=\vec x(t_0)+(t-t_0)\vec v\]
Thus, the effective coupling classical field-asymptotic current acting on a state $b^\dagger(p)\ket0$ results
\[e A^\mu_{\text{class}} j^{\text{as}}_\mu=-\frac{e}{4\pi}\frac{1}{\gamma}\frac{\delta^3\left(\vec R\right)}{\sqrt{\vec R^2+\gamma^2\left[\vec v\!\cdot\!\vec R\right]^2}}\]
Where has been used that $(n+v)^\mu(n+v)_\mu=1/\gamma^2$, Therefore, we can write
\[\kappa_i=\left(e\int A^\mu_{\text{class}} J^{\text{as}}_{\mu} dx\right)(t-t_0)\]
That is to say, These phase factors can be identified as the temporal evolution due to a Hamiltonian that comes from the coupling of the asymptotic current with one classical Lienard-Wiechert potential created by a current at a large retarded position (large at asymptotic times) and governed by the term $v_it$.

We have now to calculate the crossed phase term $\kappa_{12}$ which is given by \eqref{kappacross}
\[\kappa_{12}=\frac{e^2}{4\pi}\gamma\left(1-\vec v_1\!\cdot\!\vec v_2\right)\int_{t_0}^t ds\left\{\left[\vec x-\vec v_1t + s\, \vec v_r\right]^2+\gamma^2 \left[\vec v_1 \cdot  \left(\vec x -\vec v_1t + s \,\vec v_r\right)\right]^2\right\}^{-1/2}\]
Or, as a function of the retarded positions,
\[\kappa_{12}=\frac{e^2}{4\pi}\gamma\left(1-\vec v_1\!\cdot\!\vec v_2\right)\int_{t_0}^t ds\left\{\left[\vec R_1 + s\, \vec v_r\right]^2+\gamma^2 \left[\vec v_1 \cdot  \left(\vec R_1 + s \,\vec v_r\right)\right]^2\right\}^{-1/2}\]
We can decompose the retarded position in two components, one normal to the relative velocity and the other in its same direction
\[\vec R=\vec b + \vec R_{\parallel}\qquad \vec v_1=\vec v_{1\parallel}+\vec v_{1\bot}\]
Thus, we will obtain the following result (operating, now, with the modulus)
\[\left[\vec R_1 + s\, \vec v_r\right]^2=b^2+\left(R_\parallel+s\,v_r\right)^2 \qquad \left[\vec v_2 \cdot  \left(\vec R_1 + s \,\vec v_r\right)\right]^2=\left[v_{1\parallel} \left(\frac{v_{1\bot}}{v_{1\parallel}}b +R_\parallel +s v_r\right) \right]^2\]
For simplicity, let us also consider that $v_{1\bot}=0\Rightarrow \vec v_1=\vec v_{1\parallel}$, situation that would be satisfied, for instance, working in the center of mass frame or a frame in which either particle is at rest.
\[\kappa_{12}=\frac{e^2}{4\pi}\gamma\left(1-\vec v_1\!\cdot\!\vec v_2\right)\int_{t_0}^t ds\left\{b^2+\left( R_\parallel + s\,v_r\right)^2+\gamma^2 \left[ v_{1} \left(R_\parallel +s v_r\right) \right]^2\right\}^{-1/2}\]
performing analitically the integration we eventually obtain
\[\kappa_{12}=\frac{e^2}{4\pi}\gamma \left(1-\vec v_1\!\cdot\!\vec v_2\right) \frac{1}{\gamma v_r}\left(\ln\frac{t}{t_0}+\ln\left[\frac{\left( 1 +\dfrac{R_{\parallel}}{t\,v_r}\right)\left[1+\sqrt{\dfrac{b^2/\gamma^2}{(t\,v_r+R_\parallel)^2}+1}\right]}{\left( 1 +\dfrac{R_{\parallel}}{t_0\,v_r}\right)\left[1+\sqrt{\dfrac{b^2/\gamma^2}{(t_0\,v_r+R_\parallel)^2}+1}\right]}\right]\right)\]
For asymptotic $t,t_0$ we can neglect the quotient terms in the second logarithm according to the arguments viewed above in this section, resulting that\footnote{Although $R_\parallel$ is dominated by $-v_1t$ it can be shown that, in any case, the asymptotic behaviour of the factor $\kappa_{12}$ is correct. To see it, it is only needed to set in a frame in which $v_1=0$ and take into account that $x/t\rightarrow0$ in the asymptotic regime as we have argued above.}
\[\kappa_{12}\sim\frac{e^2}{4\pi}\gamma\left(1-\vec v_1\!\cdot\!\vec v_2\right)\frac{1}{\gamma v_r}\ln\frac{t}{t_0}\]
On the other hand, the expression of the relativistic relative velocity between two particles as a function of the 3-velocity of each one is
\[u_r=\frac{v_r}{1-\vec v_1\!\cdot\!\vec v_2}\]
So, substituting it in the previous expression we obtain
\begin{equation}\label{kappacross}\boxed{\begin{aligned}\kappa_{12}=\frac{e^2}{4\pi}\frac{1}{u_r}\ln\frac{t}{t_0}\end{aligned}}\end{equation}
Thus, the two particles state results
\begin{equation*}
\begin{split}
b_\text{as,d}^\dagger(q_1,\sigma_1)b_\text{as,d}^\dagger(q_1,\sigma_1)\ket0 &  =(2\pi)^{-6}\sum_{\sigma\sigma'}\int d^3q_1'd^3q_2'\, \sqrt{\frac{m^4}{E'_{q_1} E_{p_1'}E'_{q_2} E_{p'_2}}}u^\dagger_{\sigma}(q)u_{\sigma'}(q')u^\dagger_{\sigma}(q)u_{\sigma'}(q')\,\cdot \\
& \cdot e^{-i\kappa_1(v_1',t)}e^{-i\kappa_2(v_2',t)}\int d^3x_1\,d^3x_2\, e^{i(q_1-q_1')x_1}e^{i(q_2-q_2')x_2}e^{R(t)}e^{\chi_1(x)}e^{\chi_2(x)}e^{-i\kappa_{12}(x)}e^{i\Phi}b^\dagger(q_1') b^\dagger(q_2')\ket0
\end{split}
\end{equation*}

But, recalling the section \ref{secevol}
\[e^{i\Phi}b^\dagger(q_1') b^\dagger(q_2')\ket0=e^{i\phi}b^\dagger(q_1') b^\dagger(q_2')\ket0\]
where $\phi$ is given by \eqref{divfase} and is identically equal to $-\kappa_{12}$ and, therefore, both phase factors cancel themselves, being our state
\begin{equation*}
\begin{split}
b_\text{as,d}^\dagger(q_1,\sigma_1)b_\text{as,d}^\dagger(q_1,\sigma_1)\ket0 &  =(2\pi)^{-6}\sum_{\sigma\sigma'}\int d^3q_1'd^3q_2'\, \sqrt{\frac{m^4}{E'_{q_1} E_{p_1'}E'_{q_2} E_{p'_2}}}u^\dagger_{\sigma}(q)u_{\sigma'}(q')u^\dagger_{\sigma}(q)u_{\sigma'}(q')\,\cdot \\
& \cdot e^{-i\kappa_1(R_1,t)}e^{-i\kappa_2(R_2,t)}e^{W(t)}\int d^3x_1\,d^3x_2\, e^{i(q_1-q_1')x_1}e^{i(q_2-q_2')x_2}e^{\chi_1(x)}e^{\chi_2(x)}b^\dagger(q_1') b^\dagger(q_2')\ket0
\end{split}
\end{equation*}
where we have already applied $e^{R(t)}$ on the two particle state, and $W(t)$  (given by \eqref{vdoble}) has no fermionic operators. In order to arrive here we have used that $e^{R(t)}$ and $e^{\chi_i(x)}$ commute (It can be readily probed since the first commutator of the BCH formula vanishes)

We have now to consider the product of the operators $e^{\chi_1(x)}e^{\chi_2(x)}$ and to integrate over $d^3x_i$ and $d^3q_i'$.

It is shown in \cite{b4} that the result of the integration is (at asymptotic times)is the following \[b_\text{as,d}^\dagger(q_1,\sigma_1)b_\text{as,d}^\dagger(q_1,\sigma_1)\ket0 =e^{-i\kappa_1(R_1,t)}e^{-i\kappa_2(R_2,t)}e^{W(t)}h^{-1}_{\text{soft}}(p_1,v_1,t)h^{-1}_{\text{soft}}(p_2,v_2,t)b^\dagger(q_1,\sigma_1) b^\dagger(q_2,\sigma_2)\ket0\]
where $h^{-1}_{\text{soft}}(p_i,v_i,t)$ is called the ``minimal part'' of the dressing and it has the following expression
\[h^{-1}_{\text{soft}}(p_i,t,v_i)=\exp\left[e\int\frac{d^3k}{(2\pi)^{3/2}}\frac{1}{\sqrt{2k_0}}\left(\frac{V_i^\mu}{V_i\!\cdot\! k}e^{\frac{-i k\cdot p_i}{p_{0{_i}}}t}a_\mu(k)-\frac{V_i^\mu}{V_i\!\cdot\! k}e^{\frac{i k\cdot p_i}{p_{0{_i}}}t}a^\dagger_\mu(k)\right)\right]\]
where
\[V_i^\mu=(\eta+v_i)^\mu(\eta-v_i)\cdot k-k^\mu\]

Using the Baker Campbell Hausdorff relations we can demonstrate that
\[e^{W(p_1,p_2,t)}h^{-1}_{\text{soft}}(p_1,v_1,t)h^{-1}_{\text{soft}}(p_2,v_2,t)=e^{W'(p_1,p_2, v_1,v_2,t)}\]
where
\begin{eqnarray*}
W'(p_1,p_2, v_1,v_2,t)&\!\!\!=&\!\!\!\frac{e}{(2\pi)^{\frac32}}\int\frac{d^3k}{\sqrt{2k_0}}\left[\left(\left\{\frac{p_1^\mu}{p_1 k}-c_1^\mu\right\}e^{i\frac{k p_1}{{p_1}_{_0}}t}+\left\{\frac{p_2^\mu}{p_2 k}-c_2^\mu\right\}e^{i\frac{k p_2}{{p_2}_{_0}}t}\right)a_\mu^\dagger(k)-\right.\\
&-&\!\!\!\
\left. \left(\left\{\frac{p_1^\mu}{p_1 k}-c_1^\mu\right\}e^{-i\frac{k p_1}{{p_1}_{_0}}t}+\left\{\frac{p_2^\mu}{p_2 k}-c_2^\mu\right\}e^{-i\frac{k p_2}{{p_2}_{_0}}t}\right)a_\mu(k)\right]
\end{eqnarray*}
and where
\[c^\mu_i=\frac{V_i^\mu}{V_i\!\cdot\! k}\]
Notice that even the first commutator in the BCH formula vanishes in this case.

So we can conclude that the dressing procedure is equivalent to perform a transformation on the two particles state that operates within the asymptotic states Hilbert space since ${c^\mu_i}$ satisfy the conditions \eqref{cond1}.

Moreover, it is shown in \cite{b4} that operating on physical states on the mass shell, (that is to say, $p_i^\mu=m\gamma_i(\eta+v_i)^\mu$), the operator $\exp(W')$ is only function of the gauge fixing fields, thus, between physical states  (which are annihilated by $c$)  on the mass shell, this operator acts as the unity. We can, hence, ignore it when we are studying our case of two physical particles on the mass shell.

Eventually, we have obtained a final expression for our dressed asymptotic state of two charged particles
\begin{equation}\label{asvest}\ket{\psi_{\text{as}}^\text{d}}=e^{-i\kappa(R_1,t)}e^{-i\kappa(R_2,t)}b^{\dagger}_{\sigma_2}(p_2)b^{\dagger}_{\sigma_1}(p_1)\ket0\end{equation}
where $\kappa(R_1, t)$ and $\kappa(R_2,t)$ are given by the expressions \eqref{kappa1} and \eqref{kappa2} and they have the interpretations discussed above (evolution undergoing an effective interaction asymptotic current-classical field)

The result is completely symmetric, we could repeat the computation inverting the order of the fermionic operators and the result would be exactly the same.

\section{Entanglement of a dressed asymptotic sate of two spin $1/2$ charged fermions}

Let us build an arbitrary dressed asymptotic state of two charged particles analogously to \eqref{asvest}:
\[\ket\Psi=\sum_{\sigma_1,\sigma_2}\int d^3p_1\,d^3p_2\, \varphi_{\sigma_1\sigma_2}(p_1,p_2)\,e^{-i\kappa(p_1,t)}e^{-i\kappa(p_2,t)}\ket{\sigma_1,p_1\,;\,\sigma_2,p_2}\]
Where we are using the notation \eqref{notacion}.

The density matrix associated to this state is
\[\proj{\Psi}{\Psi}=\sum_{\sigma_1,\sigma_2}\sum_{\sigma'_1,\sigma'_2}\int d^3p_1\,d^3p_2\,d^3p'_1\,d^3p'_2\, \varphi_{\sigma_1\sigma_2}(p_1,p_2)\varphi^*_{\sigma'_1\sigma'_2}(p'_1,p'_2)\,e^{i\theta}\ket{\sigma_1,p_1\,;\,\sigma_2,p_2}\bra{\sigma'_1,p'_1\,;\,\sigma'_2,p'_2}\]
where
\[e^{i\theta}\equiv e^{-i\kappa(R_1,t)}e^{-i\kappa(R_2,t)} \,e^{i\kappa(R'_1,t)}e^{i\kappa(R'_2,t)} \]
which does not depend on the spin of the fermions. Thus, we can write the previous expression as a $4\times4$ matrix in the spin space
\begin{equation}\label{ecspin2}
\proj{\Psi}{\Psi}=\int d^3p_1\,d^3p_2\,d^3p'_1\,d^3p'_2  e^{i\theta} S_{\sigma_1\sigma_2,\sigma_1'\sigma_2'}(p_1,p_2,p'_1,p'_2) \ket{p_1,p_2}\bra{p'_1,p'_2}
\end{equation}
Being $S_{\sigma_1\sigma_2,\sigma_1'\sigma_2'}(p_1,p_2,p'_1,p'_2)$ the matrix\eqref{matriz}.

Since we will only consider spin entanglement, we have to trace over the fermions momenta
\[\int d^3q_1\,d^3q_2\braket{q_1,q_2}{\Psi}\braket{\Psi}{q_1,q_2}\]
Using that essentially $\braket{q_1,q_2}{p_1,p_2}=\bra0 b(q_1)b(q_2)b^\dagger(p_1)b^\dagger(p_2)\ket0 = \delta(q_1-p_2)\delta(q_2-p_1)$, those deltas fix $p=p'$ and, hence, given that $\vec R_i\sim -\vec v_it$, they fix $R_i'=R_i$. Therefore, in the asymptotic limit, $e^{i\theta}=1$ and all the phase terms vanish:
\[\rho_{\text{spin}}=\int d^3p_1\,d^3p_2\,S_{\sigma_1\sigma_2,\sigma_1'\sigma_2'}(p_1,p_2,p_1,p_2) \]
this expression is equal to  \eqref{rholibre}.

Thus, this is the same result which we would have obtained without considering any effect of the infrared structure, so, eventually, we have obtained that the spin entanglement is not modified by the effects derived from the presence of soft-photons

\section{Conclusions}

We have calculated the effect of the infrarred structure of the QED on the spin entanglement.

In \cite{b1}, it is pointed out the ignorance of the effect that the unavoidable presence of an undeterminate number of soft-photons could have on the entanglement of charged qubits, questioning if the very concept of qubit is no other thing that an idealization.

In order to approach this problem we have used the dressing formalism to, first of all, build physical states of two charged particles from the proper asymptotic states of QED and, afterwards, analyse what happen with their spin entanglement.

As the main result of this work, it has been demonstrated that the infrared structure of QED has no effect on the spin entanglement of charged qubits, thus all the spin correlation functions (and in particular the EPR correlations) are not modified by the presence of soft-photons when we are regarding as qubits the proper physical asymptotic states.

Finally, Along this work the next contributions have been made
\begin{itemize}
\item It has been discussed the physical relevance of the parameter $t_0$ that appears in the building of the new interaction picture.
\item It has been shown that the dressing procedure operates within the asymptotic states Hilbert that is to say $h^{-1}\mathcal{H}_{\text{as}}\subseteq \mathcal{H}_{\text{as}}$.
\item The phase contributions of the dressing procedure have been explicitly calculated for the first time.
\item It has been shown how the dressing procedure cancels the unphysical critical phase dependence on $t_0$ that comes from the construction of the asymptotic states.
\item The effect of considering the infrarred structure of the QED on a two charged fermion state has been calculated, and it has been shown that the additional terms that appear are equivalent to the temporal evolution undergoing an effective interaction Hamiltonian between the asymptotic current and the classical EM-field.
\end{itemize}

In the future we will investigate what are the effects of the infrared structure of QED regarding the momentum entanglement or the cross correlation spin-momentum and, summarizing, we will try to determine the effects of the infrarred structure of QED on all the quantum information tasks.

\newpage
\thispagestyle{plain}

\end{document}